# NGDEEP: A New Nonparametric Measure of Local Star Formation and Attenuation at Cosmic Noon

Grace M. Forrey[1,2], Raymond C. Simons[3], Jonathan R. Trump[1], Lu Shen[4,5], Anton M. Koekemoer[6], Micaela B. Bagley[7,8], Steven L. Finkelstein[7,9], Casey Papovich[4,5], and Nor Pirzkal[10]

[1] Department of Physics, 196A Auditorium Road, Unit 3046, University of Connecticut, Storrs, CT 06269, USA; gforrey@ku.edu
[2] Department of Physics and Astronomy, University of Kansas, Lawrence, KS 66045, USA
[3] Department of Engineering and Physics, Providence College, 1 Cunningham Sq., Providence, RI 02918, USA
[4] Department of Physics and Astronomy, Texas A&M University, College Station, TX, 77843-4242, USA
[5] George P. and Cynthia Woods Mitchell Institute for Fundamental Physics and Astronomy, Texas A&M University, College Station, TX 77843-4242, USA
[6] Space Telescope Science Institute, 3700 San Martin Dr., Baltimore, MD 21218, USA
[7] Department of Astronomy, The University of Texas at Austin, Austin, TX, USA
[8] Astrophysics Science Division, NASA Goddard Space Flight Center, 8800 Greenbelt Rd., Greenbelt, MD 20771, USA
[9] Cosmic Frontier Center, The University of Texas at Austin, Austin, TX, USA
[10] ESA/AURA Space Telescope Science Institute, USA

*Received 2025 December 12; revised 2026 August 14; accepted 2026 September 10; published 2026 November 10*

**Abstract**

We introduce a new nonparametric technique to quantify the spatially resolved relationship between the local star formation rate (SFR) and dust attenuation. We then apply it to 12 star-forming galaxies at $1.0 < z < 2.4$ using JWST/NIRISS slitless spectroscopy from the NGDEEP survey. First, we construct spatially resolved (∼1 kpc per pixel) Balmer decrement (H $\alpha$/H $\beta$) maps of these galaxies and derive their corresponding dust attenuation and intrinsic SFR maps. We then rank-order the map pixels by attenuation and construct a cumulative distribution curve of the total SFR as a function of increasing attenuation. We define $\mathrm{A(H\,\alpha)}^{\mathrm{SFR}}_{10\%}$, $\mathrm{A(H\,\alpha)}^{\mathrm{SFR}}_{50\%}$, and $\mathrm{A(H\,\alpha)}^{\mathrm{SFR}}_{90\%}$ as the dust attenuation levels behind which 10%, 50%, and 90% of the total integrated SFR is screened, respectively. These metrics quantify the probability that a given star-forming region lies behind a given level of attenuation. Across the full sample, 50% of the local star formation occurs behind an attenuation of 2.9 mag or higher ($\mathrm{A(H\,\alpha)}^{\mathrm{SFR}}_{50\%}$) on average. This indicates that the bulk of star formation in these galaxies is significantly attenuated by dust. The value of $\mathrm{A(H\,\alpha)}^{\mathrm{SFR}}_{10\%}$ equals 0.6 mag for the average profile, indicating that even the least attenuated star-forming regions are still highly attenuated. The globally measured attenuation more closely matches $\mathrm{A(H\,\alpha)}^{\mathrm{SFR}}_{10\%}$ than $\mathrm{A(H\,\alpha)}^{\mathrm{SFR}}_{50\%}$. This suggests that the global value is weighted toward the least dust-obscured star-forming regions and significantly underestimates the typical attenuation a star-forming region encounters. Our results demonstrate a new approach for understanding the extremely dusty local conditions of the star-forming interstellar medium in star-forming galaxies at cosmic noon.

## 1. Introduction

Dust is a key component of the interstellar medium (ISM) of galaxies despite representing only ∼1% of the ISM mass (F. Galliano et al. 2018). Dust traces cold star-forming gas, and so is closely tied with the local star formation and stellar mass growth of galaxies. Dust also helps regulate the chemical composition of the ISM. It acts as a catalyst for the creation of molecular hydrogen and as a shield from harsh UV radiation, enabling conditions for the collapse of giant molecular clouds (F. Galliano et al. 2018; H.-J. Park et al. 2024). Dust is also a nuisance, attenuating both stellar and nebular emission and posing a major challenge for interpreting observations of galaxies (S. Salim & D. Narayanan 2020). Understanding the local relationship between dust and star formation is therefore essential for determining how dust both regulates and obscures star-forming activity within galaxies (R. C. Kennicutt & N. J. Evans 2012).

The Balmer decrement, defined as the ratio of the hydrogen recombination lines H$\alpha$/H$\beta$, provides a direct probe of dust attenuation in star-forming regions. The intrinsic ratio is well understood from atomic physics under typical nebular conditions (D. E. Osterbrock & G. J. Ferland 2006). In such conditions, deviations from this intrinsic value indicate differential attenuation that preferentially suppresses H$\beta$ emission relative to H$\alpha$, due to H$\beta$ being at shorter wavelengths where stronger dust absorption occurs. Consequently, the observed Balmer decrement is routinely used to measure color excess and to infer appropriate dust corrections in both integrated and spatially resolved studies (see, e.g., S. Salim & D. Narayanan 2020 and references therein). Until recently, however, spatially resolved Balmer decrement maps were only available for nearby galaxies due to the limited sensitivity and resolution of infrared instrumentation at high redshift. Recent advances in infrared space-based instrumentation, especially with the launch of JWST (J. P. Gardner et al. 2023), have opened up new avenues for spatially resolved studies of dust attenuation and star formation in high-redshift galaxies (e.g., J. Matharu et al. 2023, 2024, 2026; P. A. Oesch et al. 2023; L. Shen et al. 2024; M. Ju et al. 2025; B. Lorenz et al. 2025; A. Tsujita et al. 2026). The epoch around redshift $z \approx 2$, commonly referred to as "cosmic noon," marks a peak in the cosmic star formation rate (SFR) and is an intense period of mass growth and structural evolution for galaxies (P. Madau & M. Dickinson 2014). Measurements of the spatially resolved characteristics of dust and SFR in galaxies at this epoch provide important insight into how galaxies build and grow during this critical period (N. M. Förster Schreiber & S. Wuyts 2020).

The slitless grism spectrographs of the Wide Field Camera 3 on the Hubble Space Telescope (HST) enabled the first statistical probes of the spatial distribution of H$\alpha$ and H$\beta$ with high resolution at these redshifts (E. J. Nelson et al. 2016; J. Matharu et al. 2022). More recently, the NIRISS and NIRCam grisms on the JWST observatory have provided a major step forward in this domain, offering slitless spectroscopy with orthogonal grisms, unprecedented sensitivity, and spatial resolution (J. Matharu et al. 2023, 2024; L. Shen et al. 2024; J. Matharu et al. 2026).

In this work, we use JWST/NIRISS slitless spectroscopy and imaging from the Next Generation Deep Extragalactic Exploratory Public (NGDEEP) survey (M. B. Bagley et al. 2024; N. Pirzkal et al. 2024) to map the Balmer decrement across 12 star-forming galaxies at $1.0 < z < 2.4$. The NGDEEP dataset allows us to study the local relationship between the dust attenuation and the SFR in these systems and to diagnose the reliability of globally derived measurements of dust attenuation. The sample is selected from the parent sample of L. Shen et al. (2024) and includes galaxies with high signal-to-noise (SNR)-integrated detections of both H$\alpha$ and H$\beta$.

We also introduce a new nonparametric analysis that quantifies the cumulative SFR as a function of the local dust attenuation acting on its light. Conceptually, this approach asks how much star formation lies behind how much dust. We argue that this approach provides a valuable new view on the link between star formation and dust attenuation on the local level.

Throughout this paper, all magnitudes are presented in the AB system (J. B. Oke & J. E. Gunn 1983; M. Fukugita et al. 1996). Where relevant, we adopt a flat $\Lambda$CDM Planck 2018 (Planck Collaboration et al. 2020) cosmology with $H_0 = 67.66\ \mathrm{km\ s^{-1}\ Mpc^{-1}}$, $\Omega_{\mathrm{m},0} = 0.30966$, and $\Omega_{\mathrm{b},0} = 0.04897$. This cosmology also assumes $T_{\mathrm{CMB}} = 2.7255$ K, $N_{\mathrm{eff}} = 3.046$, and a minimal neutrino mass sum of 0.06 eV.

## 2. Data and Sample Selection

### 2.1. Data and Processing

The galaxies studied in this paper lie in the Hubble Ultra Deep Field (HUDF; originally presented by S. V. W. Beckwith et al. 2006). They were observed with the JWST NIRISS Wide Field Slitless Spectrograph (WFSS) as a part of the NGDEEP survey (GO-2079; M. B. Bagley et al. 2024) . The WFSS dataset from NGDEEP is described in detail in M. B. Bagley et al. (2024), N. Pirzkal et al. (2024), L. Shen et al. (2024), and L. Shen et al. (2025).

The NGDEEP observations utilize the GR150R and GR150C grisms with the F115W, F150W, and F200W filters. The exposure time in F115W is 185,531 s, in F150W is 82,458 s, and in F200W is 61,844 s. The F115W observations include 36 individual exposures, and F150W and F200W observations include 12 individual exposures. Combined, these filters cover a wavelength range of 1–2.2 $\mu$m. The NGDEEP NIRISS spectroscopy reaches a 1D line flux of $\approx 1.35 \times 10^{-18}\ \mathrm{erg\ s^{-1}\ cm^{-2}}$ (L. Shen et al. 2025).

The NIRISS grism observations were taken in two position angles (PAs) in each of GR150C and GR150R grisms, providing a total of four independent dispersion PAs. This approach reduces the degeneracy between wavelength and sky position that occurs when spectra overlap along a single dispersion axis (e.g., N. Pirzkal et al. 2018). The use of two orientations enables the majority of overlapping sources to be effectively deblended, allowing for a more reliable extraction of the resolved emission line structure (e.g., P. J. Watson et al. 2025).

### 2.2. Grism Data Reduction and Extractions

The NGDEEP imaging and slitless spectroscopy was processed in L. Shen et al. (2024) and L. Shen et al. (2025) using the Grism redshift and line analysis software `Grizli` (G. Brammer 2022). `Grizli` provides a full end-to-end processing of NIRISS imaging and spectral products, including astrometric corrections, alignment, modeling, extracting, and fitting full continuum + emission line models. `Grizli` constructs emission line maps using the drizzling technique (A. S. Fruchter & R. N. Hook 2002) to combine multiple image exposures or data frames of the contamination- and continuum-subtracted 2D spectral beams back to the imaging plane. The JWST/NIRISS imaging is used to scale the data to match the total fluxes in the direct images. The emission line maps are derived using a pixel scale of $0\farcs1$. The resolved spectral energy distribution (SED) fitting and H$\alpha$ emission maps enable quantitative analysis of spatially resolved galaxy properties with a resolution of $\sim$1 (proper) kpc (L. Shen et al. 2024). The uncertainties are calculated using the drizzle weights from the constituent beam pixels. For further details on `Grizli` and its data products, we refer the reader to V. Estrada-Carpenter et al. (2019), C. Papovich et al. (2022), X. Wang et al. (2022), J. Matharu et al. (2021), R. C. Simons et al. (2021, 2023), G. Noirot et al. (2023), and J. Matharu et al. (2023).

The H$\alpha$ emission line maps are deblended from the adjacent [N II] $\lambda\lambda$6548, 6584 doublet using the mass-dependent [N II] $\lambda$6584/H$\alpha$ calibration at $z \sim 1.5$ reported by M. W. Topping et al. (2021). The weaker component of the doublet is not independently constrained by that calibration, so we fix it to $F(\lambda6548) = F(\lambda6584)/2.94$ (D. E. Osterbrock & G. J. Ferland 2006).

### 2.3. Sample Selection

The sample studied in this paper is selected from the parent sample of 19 galaxies from L. Shen et al. (2024). Galaxies in the L. Shen et al. (2024) parent sample are required to have SNR > 20 for the total Ha line flux (integrated over all pixels) and at least 50 individual pixels with Ha flux measured with SNR > 3 and were visually inspected to exclude galaxies showing contamination in one or more of their emission line maps from imperfect continuum subtraction. The `Grizli`-derived redshifts are consistent with previous ground-based and space-based spectroscopic measurements of our sources, where L. Shen et al. (2024) found a median offset of only $\Delta z/(1 + z) = 0.002$.

**Table 1**
Columns 1-4 Report the ID, R.A., decl., and Redshift of the Galaxies in Our Sample

| ID | Hα Based R.A. (deg) | Hα Based Decl. (deg) | SED Based Redshift | Corrected SFR $\log(M_\odot/yr)$ | Uncorrected SFR $\log(M_\odot/yr)$ | Stellar Mass $\log(M_\odot)$ |
|---|---|---|---|---|---|---|
| ngdeep_00503 | 53.16690 | −27.79882 | 1.992 | 2.46 ± 0.25 | 1.32 ± 0.00 | 10.70 ± 0.17 |
| ngdeep_01585 | 53.14928 | −27.78859 | 1.897 | 1.97 ± 0.20 | 1.06 ± 0.00 | 10.11 ± 0.12 |
| ngdeep_01729 | 53.16167 | −27.78750 | 1.848 | 1.93 ± 0.21 | 0.79 ± 0.00 | 10.58 ± 0.07 |
| ngdeep_02347 | 53.17432 | −27.78260 | 1.997 | 1.32 ± 0.29 | 0.77 ± 0.01 | 10.41 ± 0.08 |
| ngdeep_02703 | 53.15568 | −27.77936 | 1.844 | 1.01 ± 0.09 | 0.95 ± 0.00 | 10.56 ± 0.07 |
| ngdeep_02748 | 53.14922 | −27.77888 | 1.842 | 1.48 ± 0.27 | 0.69 ± 0.01 | 10.17 ± 0.07 |
| ngdeep_03359 | 53.17606 | −27.77377 | 1.291 | 1.17 ± 0.25 | 0.59 ± 0.00 | 10.26 ± 0.10 |
| ngdeep_03606 | 53.14618 | −27.77110 | 1.315 | 0.41 ± 0.44 | −0.60 ± 0.01 | 10.06 ± 0.10 |
| ngdeep_01524 | 53.14511 | −27.78949 | 1.302 | 0.95 ± 0.13 | 0.53 ± 0.00 | 9.83 ± 0.12 |
| ngdeep_02240 | 53.17825 | −27.78315 | 1.116 | 0.84 ± 0.19 | 0.09 ± 0.00 | 9.62 ± 0.09 |
| ngdeep_03326 | 53.14420 | −27.77362 | 1.899 | 1.58 ± 0.34 | 0.86 ± 0.00 | 9.33 ± 0.10 |
| ngdeep_03844 | 53.15231 | −27.77014 | 1.839 | 1.59 ± 0.26 | 0.87 ± 0.00 | 9.76 ± 0.09 |

**Note.** Columns (5) and (6) report the Hα-based SFR measurements (corrected and uncorrected for dust, respectively). Column (7) reports the integrated SED fit (stellar mass) from L. Shen et al. (2024).

We down-select the L. Shen et al. (2024) sample for those galaxies with both Hα and Hβ emission line coverage in order to measure the Balmer decrement. This selection imposes a redshift window of $1.0 < z < 2.4$. The final sample contains 12 galaxies, listed in Table 1.

## 3. Measurements

### 3.1. Balmer Decrement

For each galaxy, we use the Hα and Hβ maps produced by `Grizli` to derive a map of the Balmer decrement (Hα/Hβ) and its uncertainty. The Balmer decrement map and emission line maps for one of the galaxies in our sample are shown in Figure 1; the corresponding maps for the remaining 11 galaxies in the sample are provided in Figure A1 in the Appendix.

We select contiguous pixels belonging to the galaxy by constructing a segmentation map (using `photutils.segmentation`) on the Balmer decrement map. This process selects spatially connected regions with both lines detected and excludes isolated noise fluctuations. We then implement a cut on the segmentation to exclude pixels that have an Hβ SNR of <1, shown in the lower panels of Figure 1. We additionally mask two low-SNR outlier pixels (of the 720 total pixels across the sample), one each in `ngdeep_00503` and `ngdeep_01585`, with unphysically large Balmer decrements ($\mathrm{H}\,\alpha/\mathrm{H}\,\beta = 50$ and 24, corresponding to $A_V > 10$ mag and 7 mag). Both have an $\mathrm{SNR}_{\mathrm{H}\,\beta} < 1.5$ and are attributable to measurement noise rather than extreme attenuation.

We find that adopting a more stringent Hβ SNR selection of >2 leads to a systematic shift lower in the average attenuation (∼0.9 mag in the average), but does not change our qualitative results. This shift is expected, as the Balmer decrement requires a detection of Hβ, which is preferentially suppressed in the most heavily obscured regions, as discussed below.

The Balmer emission lines are produced from the recombination and subsequent "electron avalanche" to the $n = 2$ level of hydrogen atoms in the ISM. We assume case B recombination, which is appropriate for galaxies where the ISM is optically thick to Lyman series photons. Under case B recombination, the intrinsic Balmer decrement weakly depends on temperature and density. Adopting $n_e = 100$ cm$^{-3}$ and $T_e = 10^4$ K, we use $\frac{\mathrm{H}\,\alpha}{\mathrm{H}\,\beta} \approx 2.86$ (D. H. Menzel & J. G. Baker 1937; J. G. Baker & D. H. Menzel 1938; D. E. Osterbrock & G. J. Ferland 2006; W. McClymont et al. 2025). For a given total dust attenuation, the bluer Hβ line will be more attenuated than the redder Hα line. The observed $\mathrm{H}\,\alpha/\mathrm{H}\beta$ ratio thus provides a measure of the dust attenuation. We note, however, that the Balmer decrement probes only the attenuation toward regions where Hβ remains detectable, as illustrated by the contour maps in Figure 1, where Hβ is not detected in some of the brightest Hα regions. In the most heavily obscured regions the nebular emission may approach optical thickness in the Balmer lines (C. Scarlata et al. 2024), and comparisons with longer-wavelength Paschen lines indicate that Balmer-based reddening can underestimate the true nebular attenuation (N. A. Reddy et al. 2026); our inferred $A$(Hα) values are therefore conservative, a point we return to in Section 4.

As described in the next subsection, we adopt the nebular attenuation curve of N. A. Reddy et al. (2020) to derive maps of the attenuation at Hα ($A$(Hα)) and to dust-correct the Hα flux maps. The resulting maps of dust attenuation indicate attenuation that generally increases toward the centers of the galaxies. This is consistent with previous spatially resolved studies of star-forming galaxies across a range of redshifts (S. Wuyts et al. 2012; W. Wang et al. 2017; S. Hemmati et al. 2015; S. Tacchella et al. 2018).

### 3.2. Attenuation at Hα

Following I. G. Momcheva et al. (2013), we derive the attenuation at Hα by first calculating the Balmer color excess

$$E(B-V) = \frac{E(\mathrm{H}\,\beta - \mathrm{H}\,\alpha)}{\kappa(\mathrm{H}\,\beta) - \kappa(\mathrm{H}\,\alpha)} = \frac{-2.5}{\kappa(\mathrm{H}\,\beta) - \kappa(\mathrm{H}\,\alpha)} \times \log_{10}\left[\frac{(\mathrm{H}\,\alpha/\mathrm{H}\,\beta)_{\mathrm{intrinsic}}}{(\mathrm{H}\,\alpha/\mathrm{H}\,\beta)_{\mathrm{obs}}}\right] \quad (1)$$

where we assume the case B recombination value of $(\mathrm{H}\,\alpha/\mathrm{H}\,\beta)_{\mathrm{intrinsic}} = 2.86$ (D. E. Osterbrock & G. J. Ferland 2006). We adopt the quadratic form of N. A. Reddy et al.'s

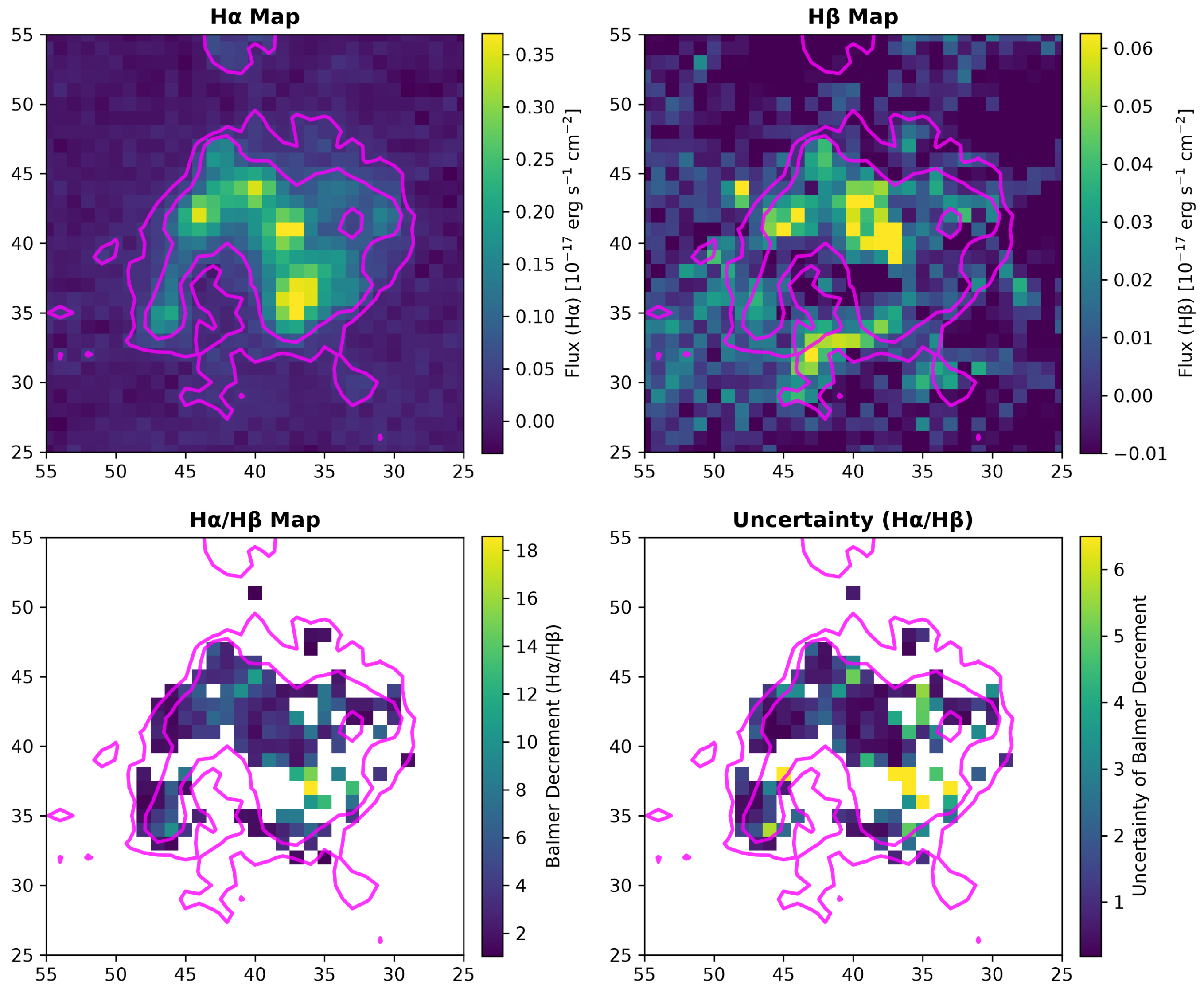

**Figure 1.** The maps derived for an example galaxy, `ngdeep_00503`, are shown with a contour map of H$\alpha$ over every panel. The top row includes the H$\alpha$ flux map (left) and the H$\beta$ flux map (right). The bottom row includes the segmented Balmer decrement (H$\alpha$/H$\beta$) map (left) and the uncertainty of the Balmer decrement map (right).

(2020) nebular attenuation curve to compute $\kappa(\lambda)$ for both H $\alpha$ and H$\beta$.

For $\kappa(\mathrm{H}\,\beta)$ and $\kappa(\mathrm{H}\,\alpha)$

$$\kappa(\lambda) = 3.719 - \frac{2.519}{\lambda} + \frac{1.196}{\lambda^2} \tag{2}$$

$$0.40 \leqslant \lambda \leqslant 0.66\ \mu\mathrm{m}$$

where $\lambda$ is $\lambda(\mathrm{H}\beta) = 0.4863\mu$m or $\lambda(\mathrm{H}\alpha) = 0.6565\ \mu$m. The attenuation at H$\alpha$ is then

$$A(\lambda) = \kappa(\mathrm{H}\,\alpha)E(B - V) \tag{3}$$

We calculate $E(B - V)$, $A(\mathrm{H}\alpha)$, and the corresponding dust-corrected SFR following the procedure described above.

### 3.3. Dust-corrected Star Formation

We derive the dust-corrected SFR of each pixel using the R. C. Kennicutt & N. J. Evans (2012) relation

$$\log_{10} \mathrm{SFR}(M_\odot\ yr^{-1}) = \log_{10} L_{\mathrm{H}\,\alpha} - 41.27 \tag{4}$$

We calculate the attenuation corrected H$\alpha$ luminosity

$$L_{\mathrm{H}\,\alpha} = f_{\mathrm{intrinsic}} \times 4\pi d_L^2 \tag{5}$$

where $f_{\mathrm{intrinsic}}(\mathrm{H}\,\alpha) = f_{obs}/10^{-0.4\times A(\mathrm{H}\alpha)}$, $f_{obs}(\mathrm{H}\alpha)$ is the observed flux at H$\alpha$ in each pixel, and $d_L$ is the luminosity distance of the galaxy determined by the redshift of the galaxy. The dust-corrected SFR is calculated for each pixel in the unmasked H$\alpha$ map. We then compute a total SFR by integrating over the segmentation map of the source.

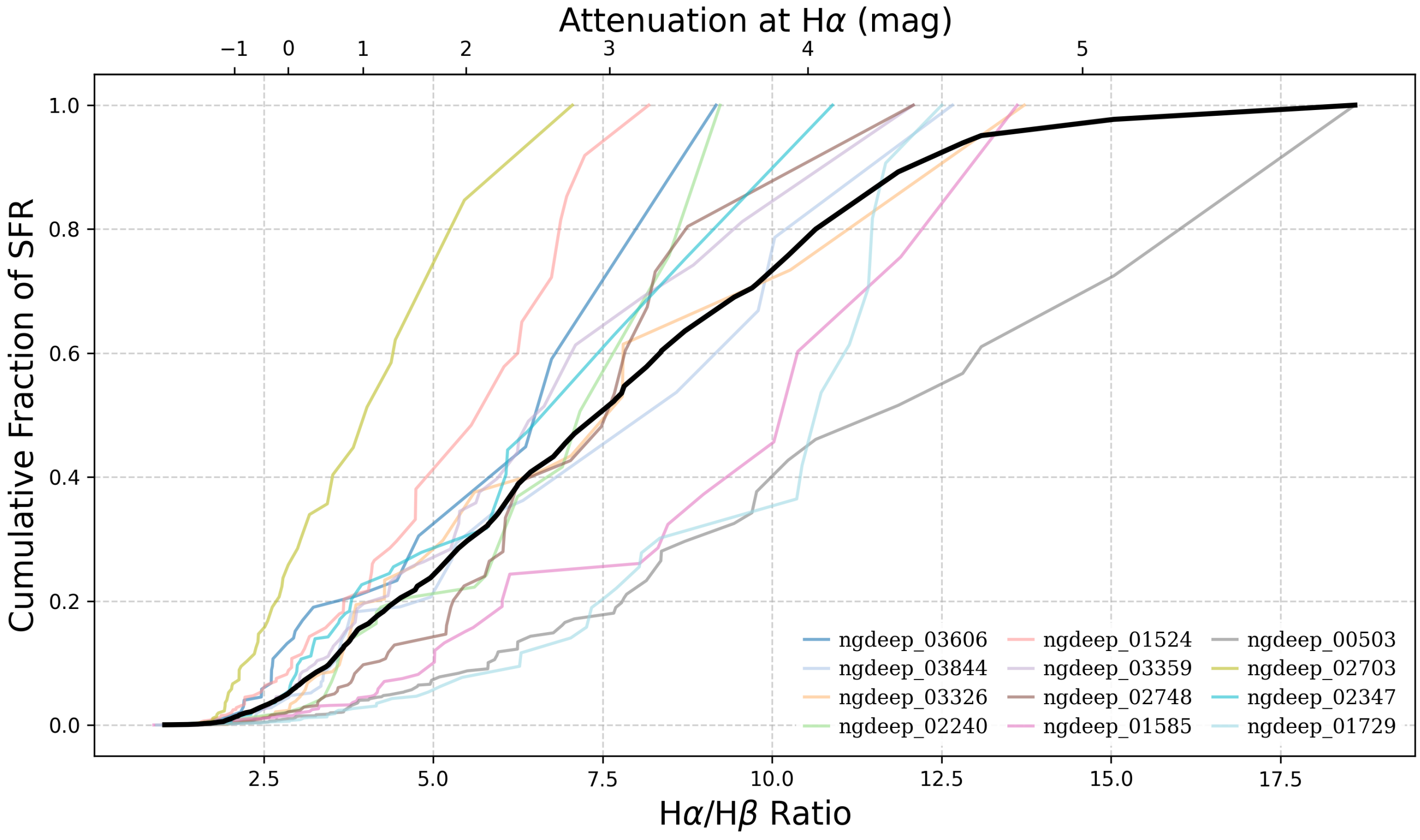

**Figure 2.** The cumulative distribution of the dust-corrected SFR as a function of the observed $H\alpha/H\beta$ emission line ratio is shown. The top axis indicates the corresponding attenuation at $H\alpha$ (in magnitudes). Conceptually, this approach quantifies the fraction of star formation at or below a given level of attenuation. The distributions for individual galaxies are shown as solid colored lines. The average distribution is shown with the bold black line.

## 4. The Cumulative Star formation Rate as a Function of Dust Attenuation

The primary goal of this paper is to assess the spatially resolved (at the pixel level) relationship between star formation and dust attenuation. Specifically, we are interested in quantifying *the fraction* of the star formation that occurs behind *increasing levels* of local dust attenuation in our sample of galaxies.

To probe that question, we introduce a new nonparametric measure: the cumulative distribution of the SFR as a function of the local Balmer decrement ($H\alpha/H\beta$). For each galaxy we reduce the 2D maps into 1D arrays sorted by the Balmer decrement of each pixel. We then cumulatively sum the SFR of the map starting from the pixel with the lowest Balmer decrement and ending at the pixel with the highest Balmer decrement. The resulting cumulative curve therefore shows the fraction of a galaxy's total star formation that is attenuated by dust at or below a given attenuation level.

In Figure 2, we show the cumulative curves for the individual galaxies in the sample (colorful lines) as well as the sample average (black bolded line). The $x$-axis is expressed in terms of the observed Balmer decrement (bottom axis) and the corresponding attenuation at $H\alpha$ ($A(H\alpha)$; top axis) using N. A. Reddy et al.'s (2020) quadratic nebular attenuation curve.

Across the full sample, 50% of the cumulative SFR corresponds to a Balmer decrement of $H\alpha/H\beta \sim 7.7$, or an attenuation at $H\alpha$ of $A(H\alpha) \sim 2.9$ mag under N. A. Reddy et al.'s (2020) nebular attenuation curve. In other words, this indicates that half of the star formation rests behind an $A(H\alpha)$ greater than 2.9 mag. This highlights the prevalence of highly obscured star formation in these galaxies. The tails of the distributions indicate the presence of star formation in highly obscured clumps, with the full sample reaching an average $H\alpha/H\beta > 10.6$ ($A(H\alpha) > 3.9$ mag).

The average curve is constructed from the full collection of SFR and Balmer decrement maps of our sample. This curve smoothly varies with increasing attenuation.

We model (and thus parametrize) the cumulative curves using the functional form

$$f(x) = A \times 0.5 \times \left[1 + \mathrm{erf}\left(\frac{x - \mu}{\sigma\sqrt{2}}\right)\right] \quad (6)$$

$$\mathrm{erf}(x) = \frac{2}{\sqrt{\pi}} \int_0^x e^{-t^2}\, dt \quad (7)$$

where $A$ is amplitude, $\mu$ determines location of the turnover, and $\sigma$ controls the width of the turnover. The erf is the standard error function. The best-fit parameters are presented in Table 2.

Curve fits with a low $A$ value correspond to galaxies with less obscured star formation overall. A fit with a low $\mu$ value indicates that most of the star formation rests in regions with low attenuation, whereas a high $\mu$ value reflects a shift toward more star formation in heavily obscured regions. Finally, fits with a small $\sigma$ indicate that the star formation is concentrated around a narrow range of attenuation, while a large $\sigma$ signifies that star formation is distributed across a broader range of attenuation. This parametrization allows us to summarize the

**Table 2**
Columns (2)–(4) List the H$\alpha$ Attenuation at 10%, 50%, and 90% of the Cumulative Star Formation Rate (A(H $\alpha)^{SFR}_{10\%}$, A(H $\alpha)^{SFR}_{50\%}$, A(H $\alpha)^{SFR}_{90\%}$)

| Galaxy ID | A(H $\alpha)^{SFR}_{10\%}$ (mag) | A(H $\alpha)^{SFR}_{50\%}$ (mag) | A(H $\alpha)^{SFR}_{90\%}$ (mag) | Global $A$(H $\alpha$) (mag) | $A$ | $\mu$ | $\sigma$ |
|---|---|---|---|---|---|---|---|
| ngdeep_01585 | 1.7 | 3.9 | 4.6 | −0.3 ± 0.1 | 1.5 | 11.9 | 3.2 |
| ngdeep_00503 | 2.2 | 4.3 | 5.5 | 0.7 ± 0.1 | 1.0 | 11.3 | 4.2 |
| ngdeep_03844 | 0.6 | 3.2 | 4.2 | 0.5 ± 0.1 | 1.0 | 7.6 | 3.1 |
| ngdeep_03326 | 0.7 | 3.0 | 4.5 | 0.7 ± 0.1 | 1.0 | 7.0 | 2.9 |
| ngdeep_02240 | 0.7 | 2.8 | 3.5 | 0.8 ± 0.2 | 3.0 | 11.0 | 3.9 |
| ngdeep_01524 | 0.1 | 2.1 | 2.8 | −0.2 ± 0.1 | 1.2 | 5.9 | 2.1 |
| ngdeep_03359 | 0.4 | 2.5 | 4.1 | 0.1 ± 0.1 | 0.9 | 6.2 | 2.3 |
| ngdeep_02748 | 1.1 | 3.0 | 4.0 | 0.5 ± 0.1 | 1.0 | 7.3 | 2.3 |
| ngdeep_02703 | −0.8 | 1.0 | 2.3 | −0.4 ± 0.1 | 1.0 | 3.9 | 1.3 |
| ngdeep_02347 | 0.2 | 2.6 | 3.8 | −0.1 ± 0.1 | 1.0 | 6.6 | 2.7 |
| ngdeep_01729 | 2.4 | 4.0 | 4.3 | 0.9 ± 0.1 | 20.5 | 23.5 | 6.7 |
| ngdeep_03606 | −0.3 | 2.5 | 3.4 | −0.1 ± 0.2 | 1.4 | 7.5 | 3.2 |
| Average Profile | 0.6 | 2.9 | 4.4 | 0.2 ± 0.0 | 0.9 | 7.2 | 3.0 |

**Note.** Column (5) reports the global $A$(H $\alpha$) derived from the integrated flux over the entire galaxy, representing the global attenuation. Columns (6)–(8) report the best-fit parameters for each galaxy using the error function model the amplitude ($A$), mean ($\mu$), and standard deviation ($\sigma$) of the fitted profile. The above quantities for the average profile are reported in the last row.

curves and compare the overall shape of attenuation profiles across our galaxy sample in a simple, quantitative way.

In Table 2 we present nonparametric measures of these curves, defined as A(H $\alpha)^{SFR}_{10\%}$, A(H $\alpha)^{SFR}_{50\%}$, and A(H $\alpha)^{SFR}_{90\%}$.

A(H $\alpha)^{SFR}_{10\%}$ is defined as the attenuation at H$\alpha$ associated with the 10$^{th}$ percentile pixel of the total integrated SFR. This value itself provides an estimate of the attenuation level associated with the *least* dusty regions of star formation. We measure an A(H $\alpha)^{SFR}_{10\%}$ of ≈0.6 mag for the average profile. This relatively high value for A(H $\alpha)^{SFR}_{10\%}$ indicates that there is very little star formation with low attenuation ($A < 1$). Similarly, A(H $\alpha)^{SFR}_{50\%}$ corresponds to the attenuation value of the 50$^{th}$ percentile pixel of SFR. In other words, half of the integrated SFR will be attenuated at a greater level than A(H $\alpha)^{SFR}_{50\%}$ and half will be attenuated at a lower level. This value can be considered a *typical* or SFR-weighted median attenuation in the galaxy. For the average profile, the value of A(H $\alpha)^{SFR}_{50\%}$ is ≈2.9 mag. Half of the SFR is attenuated at greater than 2.9 mag, again highlighting the extremely dusty conditions in these galaxies. Lastly, A(H $\alpha)^{SFR}_{90\%}$ corresponds to the attenuation of the 90$^{th}$ percentile of the integrated SFR. At A(H $\alpha)^{SFR}_{90\%}$, we measure the attenuation of the average profile as ≈ 4.4 mag. Importantly, the metrics A(H $\alpha)^{SFR}_{10\%}$, A(H $\alpha)^{SFR}_{50\%}$, and A(H $\alpha)^{SFR}_{90\%}$ are percentiles read directly from the empirical, nonparametric cumulative SFR attenuation distribution. They do not rely on the error function parametrization, which we provide only as a compact summary of curve shape.

We now focus our attention on the similarities and differences in the individual profiles. We do not observe a strong dependence between a galaxy's position in Figure 2 and whether it is above or below the average attenuation profile. Although all galaxies show a similar linear increase in attenuation from 10% to 90% cumulative SFR, there is significant galaxy-to-galaxy variation in both slope and amplitude of the curves. These results highlight the diversity in dust distribution and geometry across the sample.

To help interpret these cumulative profiles, it is helpful to consider how toy dust geometries would present in Figure 2. If dust is distributed as a uniform screen with constant optical depth across the galaxy, all star-forming regions would be equally attenuated. In this case, the curves would sharply rise as a step function at a single attenuation value. This contrasts the smoother, more gradual increase observed in Figure 2. Real galaxies have spatially varying optical depths and cannot be described by a simple uniform foreground screen with a constant optical depth. Another useful case is if dust is uniformly distributed in the galaxy. In this situation, star-forming regions that reside further in would be more obstructed than those at the surface. As a result, the cumulative profile would rise smoothly and linearly. Aside from their tails, the curves in Figure 2 all show a characteristic linear behavior. This is consistent with the expectations described above for uniform dust.

In Table 2, we also list the global $A$(H$\alpha$) values for each galaxy. These are derived by summing all pixels together from the segmented H$\alpha$ and H$\beta$ maps and calculating a single (global) Balmer decrement and attenuation at H$\alpha$.

Two galaxies, `ngdeep_03606` and `ngdeep_02703`, exhibit nearly zero attenuation at A(H $\alpha)^{SFR}_{10\%}$ and at global $A$ (H$\alpha$). The negative $A$(H$\alpha$) values in Table 2 arise from their unphysical Balmer decrement measurements: in 71% of `ngdeep_02703`'s pixels and 57% of `ngdeep_03606`'s pixels the measured $\frac{H\,\alpha}{H\,\beta}$ falls below the intrinsic case B value of 2.86. This could be due to minor errors in the data processing (e.g., oversubtracted backgrounds) that lead to artificially boosted H$\beta$ or reduced H$\alpha$ emission. Or it could be due to a departure from case B, perhaps due to gas that is optically thick to Balmer photons (C. Scarlata et al. 2024). In reality, these galaxies are probably nearly dust free, with a Balmer decrement that is close to the intrinsic value. The negative attenuation values therefore reflect processing or measurement effects rather than actual negative attenuation.

Naively, the spatially integrated $A$(H$\alpha$) should be comparable to A(H $\alpha)^{SFR}_{50\%}$, representing the dust attenuation experienced by the median star-forming regions within a galaxy. This expectation holds for the observed flux: integrating the spaxel fluxes reproduces the integrated flux. It does not hold for the dust-corrected SFR because the dust correction is a nonlinear function ($10^{0.4\,A(\mathrm{H}\,\alpha)_i}$) of the local attenuation. Correcting each spaxel and then summing

$$\mathrm{SFR}_{\mathrm{resolved}} \propto \sum_i f_{\mathrm{obs},i}\; 10^{0.4\,A(\mathrm{H}\,\alpha)_i}, \quad (8)$$

is therefore not equivalent to summing the observed flux and applying a single global correction

$$\mathrm{SFR}_{\mathrm{global}} \propto \left(\sum_i f_{\mathrm{obs},i}\right) 10^{0.4\,A(\mathrm{H}\,\alpha)_{\mathrm{global}}}. \quad (9)$$

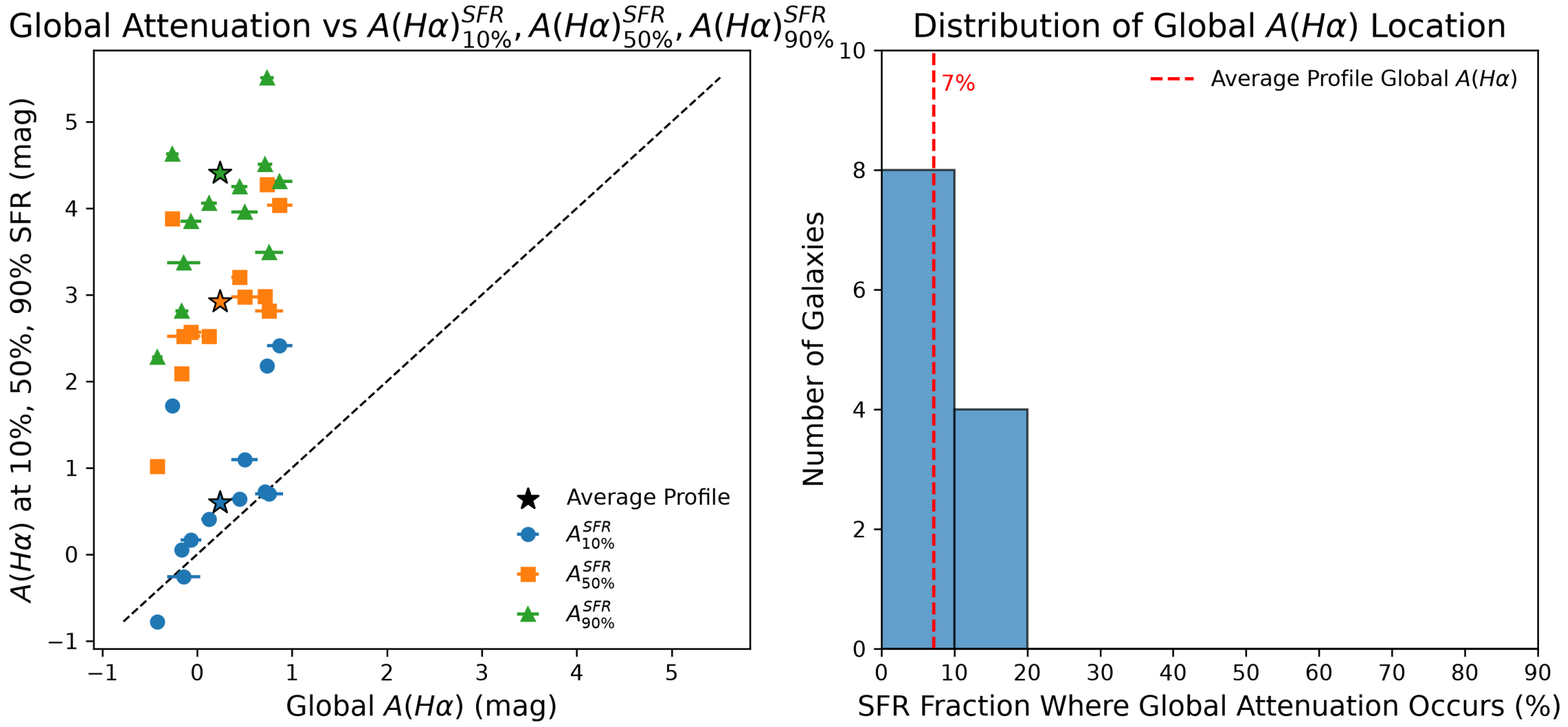

**Figure 3.** The left panel shows the global $A(\mathrm{H}\alpha)$ versus the quantities $\mathrm{A(H\,\alpha)}^{\mathrm{SFR}}_{10\%}$ (blue circle), $\mathrm{A(H\,\alpha)}^{\mathrm{SFR}}_{50\%}$ (orange square), and $\mathrm{A(H\,\alpha)}^{\mathrm{SFR}}_{90\%}$ (green triangle). The stars with black outlines represent the corresponding values for the average profile (see Figure 2). The right panel shows the cumulative SFR fraction corresponding to the attenuation measured globally. In short, this distribution indicates the fraction of new stars forming at or below the global attenuation level. The red dashed line indicates the SFR fraction associated with the global $A(\mathrm{H}\alpha)$ measured for the average profile.

The two agree only if the attenuation is spatially uniform. For any spatially varying attenuation, the resolved sum will differ from the globally corrected value. The difference between the two is itself a measure of the inhomogeneity of the attenuation.

A comparison of the spatially integrated $A(\mathrm{H}\alpha)$ values in Table 2 with the cumulative profiles reveals that the integrated attenuation is most consistent with $\mathrm{A(H\,\alpha)}^{\mathrm{SFR}}_{10\%}$.

In the left panel of Figure 3, we compare the spatially integrated measurement of $A(\mathrm{H}\alpha)$ with the attenuation value corresponding to the 10th, 50th, and 90th percentile of SFR. Here we confirm that nearly all galaxies in our sample show a close relationship between the global $A(\mathrm{H}\alpha)$ and $\mathrm{A(H\,\alpha)}^{\mathrm{SFR}}_{10\%}$.

In the right panel of Figure 3, we quantify the cumulative percentile associated with the global attenuation value. This panel shows the distribution of the percentile across the sample. In other words, this panel shows the fraction of the total star formation that occurs at or below the globally measured $A(\mathrm{H}\alpha)$. The histogram is strongly skewed toward low cumulative fractions, with all galaxies having their global attenuation correspond to only ∼0%–20% of the integrated SFR. For the average profile, the global $A(\mathrm{H}\alpha)$ aligns with 7% of the cumulative SFR, indicated by the red dashed line. This demonstrates that the global attenuation primarily reflects the least obscured, most luminous regions that dominate the total H α flux rather than the dust conditions affecting the bulk of the star formation. In other words, the integrated $A(\mathrm{H}\alpha)$ systematically samples the low-attenuation tail of the distribution rather than the SFR-weighted median. Global measurements significantly underestimate the true level of obscuration experienced by most star formation within these galaxies. A similar effect is reported by A. Tsujita et al. (2026) at $z \sim 4$–6, where spatially integrated estimates of dust attenuation fall systematically below the corresponding pixel-by-pixel measurements, consistent with integrated light being dominated by the least obscured sightlines.

We note that Balmer decrements may themselves underestimate the true nebular reddening. Using multiple Balmer and Paschen lines from the JWST/AURORA survey, N. A. Reddy et al. (2026) derived nebular attenuation curves and dust covering fractions for 24 star-forming galaxies at $1.5 < z < 4.4$, and found that at least half of their sample showed greater reddening from Paschen lines than from Balmer lines. Should similar conditions hold in our sample, our attenuation values would be lower bounds, and the true fraction of star formation occurring behind high attenuation could be larger than we measure.

This difference underscores the value of the cumulative attenuation analysis presented here, which provides a more nuanced view of dust obscuration across the whole galaxy. The cumulative profiles are derived on a pixel-by-pixel basis, tracing how attenuation varies across spatially distinct star-forming regions. In contrast, the global $A(\mathrm{H}\alpha)$ compresses all regions into a single (light-weighted) average. As a result, the global attenuation often underestimates the true level of obscuration, since the global attenuation effectively only captures <10% of the total SFR leaving >90% of a galaxy's SFR unaccounted for. In this sense, relying solely on integrated measurements could lead to substantial underestimates of the total SFR and misrepresent the diversity of dust geometries within galaxies. The cumulative approach, by contrast, provides a more complete and physically meaningful characterization of internal attenuation and its impact on star formation.

Overall, the cumulative profile provides a powerful and physically motivated framework for comparison with theoretical models of dust geometry and radiative transfer.

## 5. Conclusions

We introduce a new nonparametric technique to study the pixel-level relationship between dust attenuation and star formation. We apply this technique on NIRISS slitless spectroscopy of 12 galaxies at $1.0 < z < 2.4$ acquired as a part of the NGDEEP survey (M. B. Bagley et al. 2024). The primary conclusions of this paper are as follows:

1. Across the full sample, 50% of the local star formation occurs behind an attenuation of ∼2.9 mag or higher ($A(\mathrm{H}\,\alpha)^{\mathrm{SFR}}_{50\%}$) on average, indicating that the bulk of star formation in these galaxies is significantly attenuated by dust.
2. A nonnegligible fraction (10%) of star formation is associated with extreme attenuation (>3.9 mag of attenuation).
3. Global $A(\mathrm{H}\alpha)$ values underestimate the *typical* obscuration of individual star-forming regions. The average global $A(\mathrm{H}\alpha)$ corresponds to the SFR percentile of 7%. This indicates that the global value is biased by the brightest, least obscured regions. The global measurement fails to capture the true high attenuation impacting the light of the bulk of newly forming stars.

Our results demonstrate the power of JWST slitless spectroscopy, combined with our new nonparametric measure of local star formation and attenuation, to probe internal dust and star formation structures in galaxies at the peak of cosmic activity. The measurement introduced here allows us to directly quantify how much star formation is distributed across different levels of obscuration, providing a physically intuitive, probabilistic description of the dusty conditions experienced by star-forming regions within individual galaxies. In doing so, it offers a fundamentally new view compared to traditionally used global (attenuated) measurements. Our results reveal obscuration that would otherwise be underestimated by these traditional measurements. As JWST continues to expand the frontier of high-resolution infrared spectroscopy, future studies incorporating larger, more diverse samples, especially those probing lower mass systems and extending to even higher redshifts, will refine our understanding of the processes governing galaxy growth and star formation during this pivotal epoch in cosmic history.

## Acknowledgments

G.F. and R.C.S. acknowledge support from the NASA Connecticut Space grant Consortium. We thank the NGDEEP collaboration for their dedication and for providing the datasets used in this work. We thank the anonymous referee for a careful and constructive report that has improved both the clarity and the analysis presented in this paper. We also thank the AAS Statistics Editor for helpful comments on this work. This research made use of observations and data obtained through the NGDEEP program from JWST.

The JWST data presented in this article were obtained from the Mikulski Archive for Space Telescopes (MAST) at the Space Telescope Science Institute. The specific observations analyzed can be accessed via doi:10.17909/k2hc-7n64.

*Facility:* JWST.

## Appendix

Balmer decrement and emission line maps for 11 of the 12 galaxies in our sample, the maps for the remaining galaxy are shown in Figure A1.

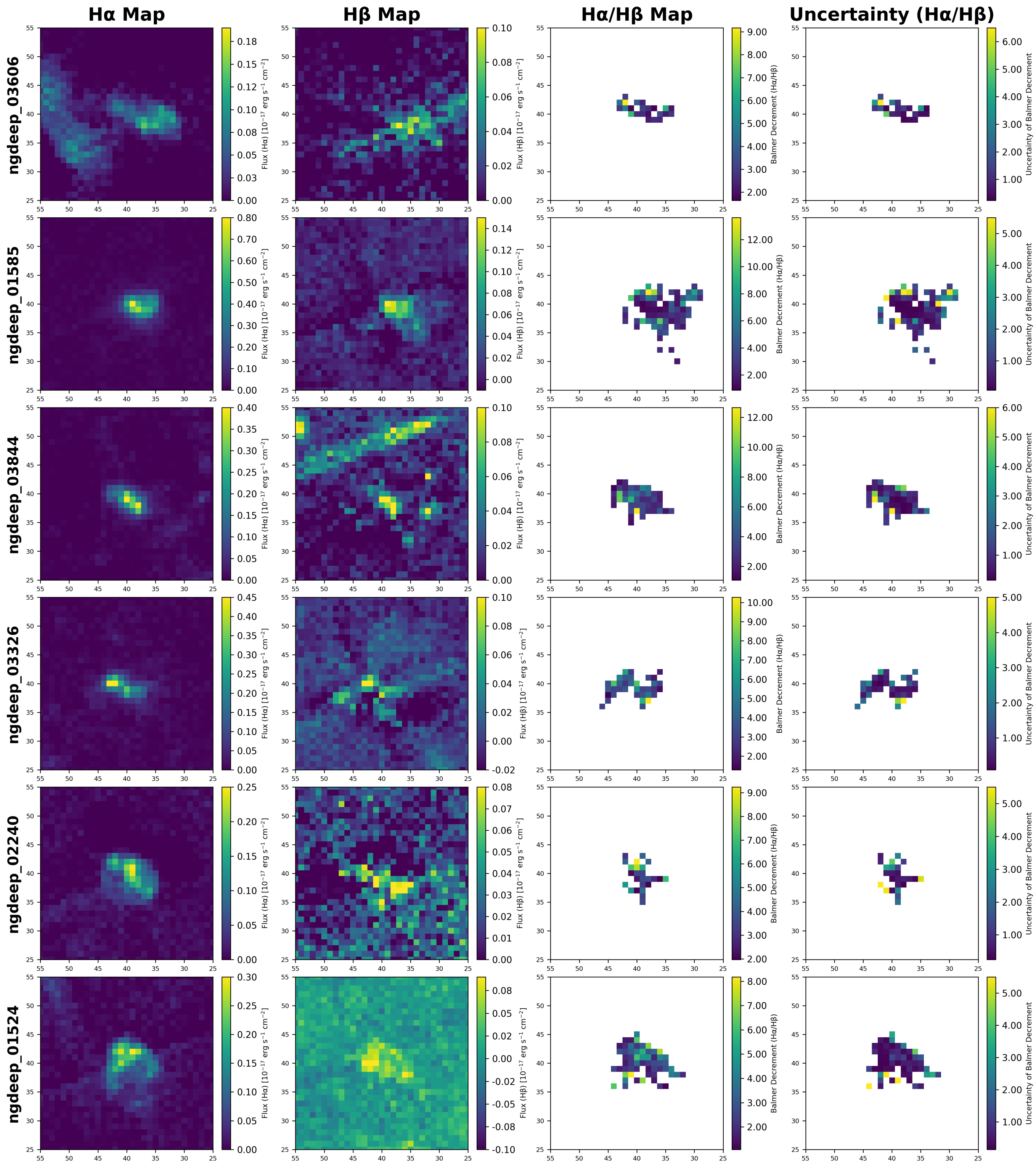

**Figure A1.** Emission line and Balmer decrement maps for the galaxies in our sample (excluding `ngdeep_00503`, which is shown in Figure 1). Each row shows an individual galaxy. From left to right: the H$\alpha$ flux map, the H$\beta$ flux map, the segmented Balmer decrement (H$\alpha$/H$\beta$) map, and the uncertainty of the Balmer decrement map.

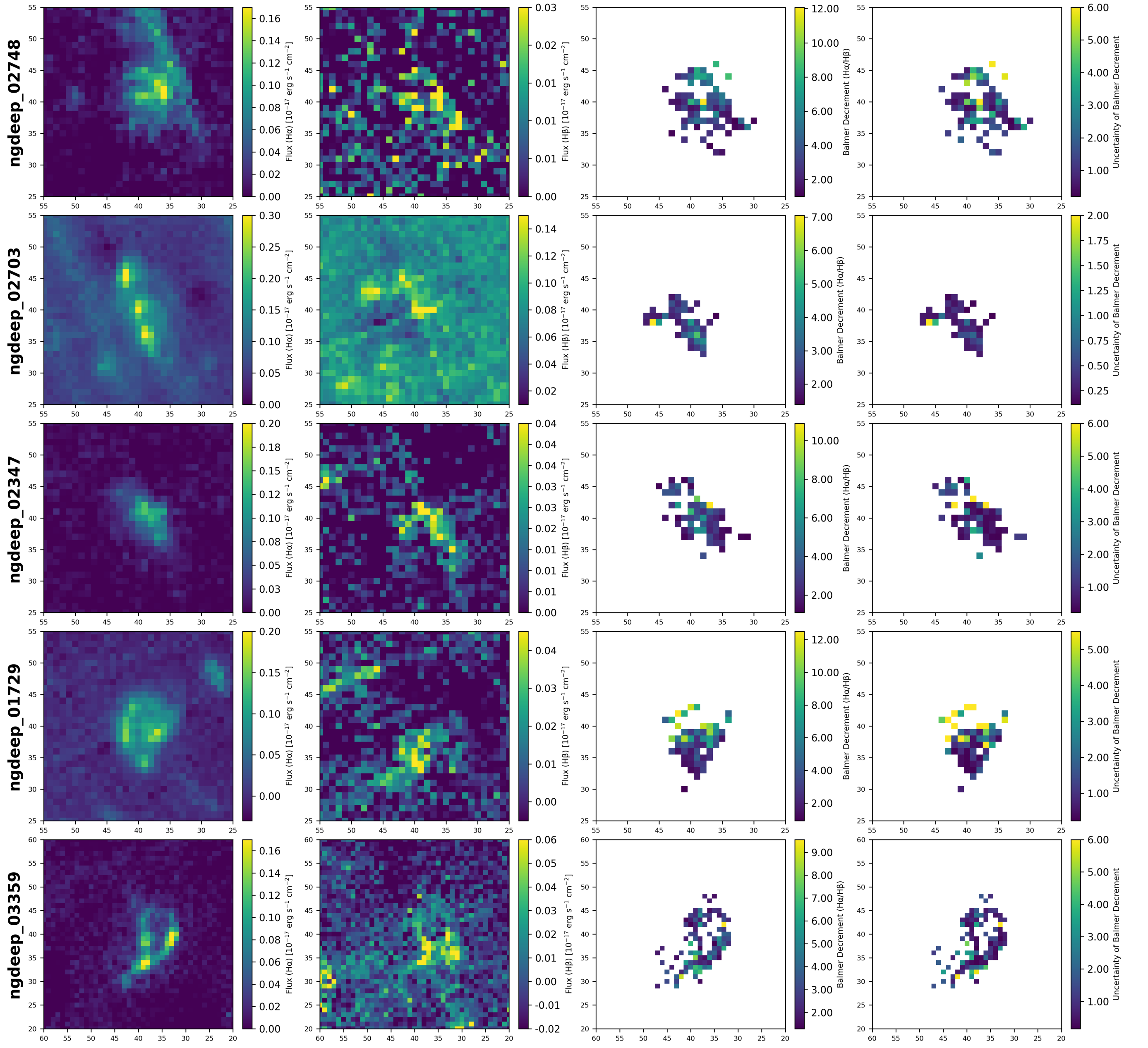

**Figure A1.** (Continued.)

## ORCID iDs

Grace M. Forrey https://orcid.org/0009-0003-5179-6942
Raymond C. Simons https://orcid.org/0000-0002-6386-7299
Jonathan R. Trump https://orcid.org/0000-0002-1410-0470
Lu Shen https://orcid.org/0000-0001-9495-7759
Anton M. Koekemoer https://orcid.org/0000-0002-6610-2048
Micaela B. Bagley https://orcid.org/0000-0002-9921-9218
Steven L. Finkelstein https://orcid.org/0000-0001-8519-1130
Casey Papovich https://orcid.org/0000-0001-7503-8482
Nor Pirzkal https://orcid.org/0000-0003-3382-5941